\documentclass[11pt]{article}
\usepackage{amsfonts}
\usepackage{amssymb}
\usepackage{amsmath}
\usepackage{graphicx}
\usepackage{latexsym}

\newcommand{\op}[1]{\operatorname{#1}}

\def \qed {\hfill \rule{2mm}{2mm}\vspace{3mm}}

\newtheorem{theorem}{Theorem}
\newtheorem{lemma}[theorem]{Lemma}

\newtheorem{prop}[theorem]{Proposition}

\newtheorem{fact}[theorem]{Fact}

\begin{document}

\title{\Large\bf Continuous-time quantum walks on the symmetric group}

\author{
Heath Gerhardt \hspace{1.5cm} John Watrous\\[2mm]
Department of Computer Science\\
University of Calgary\\
Calgary, Alberta, Canada\\
\{gerhardt,\,jwatrous\}@cpsc.ucalgary.ca
}

\date{May 29, 2003}

\maketitle

\begin{abstract}
In this paper we study continuous-time quantum walks on Cayley graphs of the
symmetric group, and prove various facts concerning such walks that
demonstrate significant differences from their classical analogues.
In particular, we show that for several natural choices for generating sets,
these quantum walks do not have uniform limiting distributions, and are
effectively blind to large areas of the graphs due to destructive interference.
\end{abstract}


\section{Introduction}

According to our current understanding of physics, quantum mechanics provides
sources of true randomness, and mathematically speaking much of the underlying
framework of quantum information and computation may be viewed as an extension
of the study of random processes.
The focus in quantum information and computation is often placed on finding
information processing tasks that can be performed with the help of
quantum information (such as factoring integers in polynomial
time~\cite{Shor97} or implementing unconditionally secure key
distribution~\cite{BennettB84,ShorP00}) or on studying the distinctively
non-classical aspects of quantum information (such as entanglement; see,
for instance,~\cite{HorodeckiH+01}).
However, it seems quite plausible that the study of quantum information and
computation will also lead to new methods in the study of classical
computation and random processes.
Along these lines, Kerenidis and de Wolf~\cite{KerenidisW03} recently used
quantum arguments to prove new results on (classical) locally decodable codes.

As a step toward understanding the possible implications of quantum
methods for the study of random processes, it is natural to consider the
differences between classical and quantum processes.
One of the topics that has recently received attention in the quantum
computing community that highlights these differences is the the study of
quantum computational variants of random walks, or {\em quantum walks}
\cite{AharonovA+01,AhmadiB+03,AmbainisB+01,BachC+02,ChildsC+03,ChildsF+02,
FarhiG98,Kempe02,MackayB+01,MooreR02,YamasakiH+02}.
(A recent survey on quantum walks by Kempe~\cite{Kempe03} is an ideal
starting point for background on quantum walks.)
In this paper we consider quantum walks on Cayley graphs of the
symmetric group---a topic that has been suggested in at least two previous
papers on quantum walks~\cite{Kempe03,AhmadiB+03}.

Two main variants of quantum walks have been considered: continuous-time
quantum walks and discrete-time quantum walks.
We restrict our attention to continuous-time quantum walks in this paper.
Keeping in line with previous results on quantum walks, we find some
significant differences between quantum and classical random walks on
Cayley graphs of the symmetric group.
In particular, we find that quantum walks on Cayley graphs of the
symmetric group do not have uniform limiting distributions for several
natural choices for the generators.
This answers a question recently suggested by Ahmadi, Belk, Tamon, and
Wendler~\cite{AhmadiB+03} concerning non-uniform mixing of quantum walks.

One of the principle motivations for studying quantum walks has been that
quantum walks may potentially be useful as algorithmic tools.
This potential was recently demonstrated by Childs, Cleve, Deotto, Farhi,
Gutmann and Spielman \cite{ChildsC+03}, who prove that there exists a black-box
problem for which a quantum algorithm based on quantum walks gives an
exponential speed-up over any classical randomized algorithm.
The key to this algorithm is that a quantum walk is able to permeate a
particular graph while any classical random walk (or any classical randomized
algorithm, for that matter) cannot.
One of the first problems that comes to mind as an obvious challenge for the
quantum algorithms community is the graph isomorphism problem, and
it is natural to ask whether quantum walks, and in particular quantum
walks on Cayley graphs of the symmetric group, can be of any use for an
algorithm for this problem.
(While this was our primary motivation for studying quantum walks on the
symmetric group, we have not found any way to apply our results to
this problem.)


\section{Definitions}

\subsection{Continuous-time quantum walks on graphs}

\label{sec:walks-definition}

A continuous-time quantum walk on an undirected graph $\Gamma = (V,E)$ can
be defined in the following way.
First, we let $A$ be the $|V|\times |V|$ adjacency matrix of $\Gamma$,
i.e., $A$ is indexed by elements of $V$ and is as follows:
\[
A[u,v] = \left\{
\begin{array}{ll}
1 & \mbox{if $\{u,v\}\in E$}\\
0 & \mbox{otherwise}
\end{array}
\right.
\]
Next, let $D$ be the $|V|\times |V|$ diagonal matrix for which the diagonal
entry corresponding to vertex $v$ is $\op{deg}(v)$ and let $L = D - A$.
The matrix $L$ is positive semidefinite and, under the assumption that
$\Gamma$ is connected, 0 is an eigenvalue with multiplicity 1; the uniform
vector is a corresponding eigenvector.
The quantum walk on $\Gamma$ is then given by the following unitary
matrix:
\[
U(t) = e^{-it L}
\]
for $t\in\mathbb{R}$.
If the quantum walk on $\Gamma$ is run for time $t$ starting at vertex $u$,
then the amplitude associated with each vertex $v$ is $U(t)[v,u]$, and thus
measuring at this point (with respect to the standard basis) results in each
vertex $v$ with probability $\left|U(t)[v,u]\right|^2$.
If instead of starting at a particular vertex $u$ we have some quantum state
described by $\psi:V\rightarrow\mathbb{C}$, and we run the quantum walk for
time $t$, the new quantum state is described by $U(t)\psi$, and measuring
results in each vertex $v$ with probability $\left|(U(t)\psi)[v]\right|^2$.
Other types of measurements can be considered, but we will focus just
on this sort of measurement where the outcome is a vertex of the graph. 
To our knowledge, continuous-time quantum walks were first considered
by Farhi and Gutmann~\cite{FarhiG98}.

Continuous-time quantum walks are analogous to continuous-time random
walks on $\Gamma$, where the evolution is described by
\[
M(t) = e^{-tL}
\]
rather than $U(t)$ as above.
Specifically, if the continuous-time random walk is started at vertex
$u$ and run for time $t$, the probability of being at vertex $v$ is
given by $M(t)[v,u]$.
Continuous-time random walks share many properties with their discrete-time
variants \cite{AldousF02}.

This paper is concerned with quantum walks on Cayley graphs, which are
regular graphs.
In the case of regular graphs there is no difference between using the
matrix $L$ and the adjacency matrix for the definition of quantum walks, and
we find it is more convenient to use the adjacency matrix for the
graphs we are considering.
(Of course one cannot replace $L$ with the adjacency matrix
when discussing the classical case, since this would not give rise
to a stochastic process---the equivalence only holds for the quantum
case.)
The reasoning behind this equivalence is as follows.
Because $D$ and $A$ commute for regular graphs, we see that
\[
U(t) = e^{-it d I} e^{it A} = e^{-itd} e^{it A};
\]
the difference is a global phase factor, which has no significance when
calculating the probabilities.
So, from here after in this paper we will consider the unitary
process given by $U(t) = e^{it A}$ rather than $e^{-itL}$.

In the case of classical random walks, there are various properties of
random walks that are of interest.
One of the most basic properties of a classical random walk is
the limiting distribution (or stationary distribution).
This distribution is the uniform distribution for random walks on connected,
regular graphs, and in fact as a result of the way we have defined
continuous-time random walks this distribution is uniform for any connected,
undirected graph;
this is apparent by considering the spectral decomposition of
the matrix~$e^{-tL}$.

As quantum walks are unitary (and therefore invertible) processes,
they do not converge to any state, so one must be precise about what is
meant by the limiting distribution.
Suppose we have a quantum walk on some graph $\Gamma$ and some vertex
$u$ has been designated as the starting vertex.
The probability of measuring the walk at some vertex $v$ after time
$t$ is, as described above, given by
\[
P_t[v] = \left|U(t)[v,u]\right|^2.
\]
If $t$ is chosen uniformly from some range $[0,T]$ then the resulting
distribution is
\[
\bar{P}_T[v]= \frac{1}{T}\int^T_0 P_t[v]\mathrm{d}t.
\]
In the limit for large $T$ these distributions converge to some
distribution $\bar{P}$, which is the limiting distribution of the
quantum walk.
This notion of the limiting distribution for a quantum walk is discussed
in \cite{AharonovA+01}.


\subsection{Cayley graphs and representation theory of the symmetric group}
\label{sec:definitions-walks}

In this section we briefly discuss necessary background information
on Cayley graphs of the symmetric group and on representation theory
of the symmetric group, which is the main tool used in this paper
to analyze quantum walks on Cayley graphs.

Let $G$ be a finite group and let $R\subseteq G$ be a set of generators
for $G$ satisfying $g\in R\Leftrightarrow g^{-1}\in R$ for all $g\in G$.
Then the Cayley graph of $G$ with respect to $R$, which we denote by
$\Gamma(G,R)$ in this paper, is an undirected graph defined as follows.
The set of vertices of $\Gamma(G,R)$ coincides with $G$, and for any
$g,h\in G$, $\{g,h\}$ is an edge in $\Gamma(G,R)$ if and only if
$g h^{-1}\in R$.
Equivalently, if $R = \{h_1,\ldots,h_d\}$ then each vertex $g$ is adjacent to
vertices $h_1 g,\ldots,h_d g$.
Thus, $\Gamma(G,R)$ is a regular graph of degree $d = |R|$.
We will restrict our attention to generating sets that form conjugacy classes.
(The method we use for analyzing quantum walks on Cayley graphs is limited to
such generating sets.)
Recall that for some group $G$, elements $g$ and $h$ are conjugate if there
exists some $a\in G$ such that $a^{-1}g a = h$.
This is an equivalence relation that partitions $G$ into conjugacy classes.
A function $f:G\rightarrow\mathbb{C}$ is a {\em class function} if it
is constant on conjugacy classes of $G$.

The conjugacy classes in $S_n$ are determined by the cycle structures
of elements when they are expressed in the usual cycle notation.
Recall that a partition $\lambda$ of $n$ is a sequence
$(\lambda_1,\ldots,\lambda_k)$ where $\lambda_1\geq\cdots\geq\lambda_k\geq 1$
and $\lambda_1+\cdots+\lambda_k = n$.
The notation $\lambda\vdash n$ indicates that $\lambda$ is a partition of $n$.
There is one conjugacy class for each partition $\lambda \vdash n$ in
$S_n$, which consists of those permutations having cycle structure described
by $\lambda$.
We denote by $C_{\lambda}$ the conjugacy class of $S_n$ consisting of all
permutations having cycle structure described by~$\lambda$.

A {\em representation} of a group $G$ is a homomorphism from $G$ to
$\op{GL}(d,\mathbb{C})$ for some positive integer $d$, where
$\op{GL}(d,\mathbb{C})$ denotes the general linear group of invertible
$d\times d$ complex matrices.
The {\em dimension} of such a representation is $d$, and we write
$\op{dim}(\rho)$ to denote the dimension of a given representation $\rho$.
Two representations $\rho_1:G\rightarrow\op{GL}(d_1,\mathbb{C})$ and
$\rho_2:G\rightarrow\op{GL}(d_2,\mathbb{C})$ are {\em equivalent} if there
exists an invertible linear mapping
$A:\mathbb{C}^{d_1}\rightarrow\mathbb{C}^{d_2}$ such that
$A\rho_1(g) = \rho_2(g)A$ for all $g\in G$, otherwise they are
{\em inequivalent}.
A representation $\rho$ of dimension $d$ is {\em irreducible} if there are no
non-trivial invariant subspaces of $\mathbb{C}^d$ under $\rho$.
That is, if $W \subseteq\mathbb{C}^d$ is a subspace of $\mathbb{C}^d$
such that $\rho(g)W \subseteq W$ for all $g\in G$, then
$W = \mathbb{C}^d$ or $W = \{0\}$.
A collection of inequivalent, irreducible representations is said to be
{\em complete} if every irreducible representation is equivalent to
one of the representations in this set.
It holds that any complete set of irreducible representations can
be put into one-to-one correspondence with the conjugacy classes
of the group in question.

The character corresponding to a representation $\rho$ is a mapping
$\chi_\rho:G\rightarrow\mathbb{C}$ obtained by taking the trace of the
representation: $\chi_\rho(g) = \op{tr}(\rho(g))$.
Using the cyclic property of the trace it follows that the characters are
constant on the conjugacy classes of a group.
If we have a complete set of inequivalent, irreducible representations
of a group, then the corresponding characters form an orthogonal basis for the
space of all class functions.

The Fourier transform $\hat{f}$ of a complex-valued function $f$ on $G$
at a representation $\rho$ is
\[
\hat{f}(\rho) = \sum_{g\in G}f(g)\rho(g).
\]
\begin{fact}
\label{fact:Fourier-class}
Let $f$ be a class function on a group $G$ and $\rho$ be an irreducible
representation of $G$, then $$\hat{f}(\rho) = 
\frac{1}{\op{dim}(\rho)}\left(\sum_{g\in G}f(g)\chi_\rho(g)\right)I$$
\end{fact}

For the symmetric group on $n$ elements there is a particular way of
associating the partitions of $n$ (which are in one-to-one correspondence
with the conjugacy classes of $S_n$) with a complete set of inequivalent,
irreducible representations of $S_n$.
These particular representations are said to be in {\em Young normal form}.
(Several text books describe the specific method for constructing these
representations---see, for instance, James and Kerber~\cite{JamesK81}.
It will not be important for this paper to discuss the actual construction
of these representations.)
These representations have the special property that all matrix entries
in these representations are integers.
Once we have these irreducible representations, it is possible to associate
with each one an equivalent irreducible representation that has the property
that $\rho(g)$ is a unitary matrix for every $g\in S_n$.
The irreducible, unitary representation associated with a given partition
$\lambda\vdash n$ will be denoted $\rho_\lambda$, and the corresponding
character will be denoted $\chi_\lambda$.
The following fact will be a useful fact regarding these representations.

\begin{fact}
\label{fact:orthogonality-relations}
Let $\lambda$ and $\mu$ be partitions of $n$ and let
$\rho_\lambda$ and $\rho_\mu$ be the associated unitary representations
as described above.
Then for all $1\leq i,j\leq\op{dim}(\rho_\lambda)$
and $1\leq k,l\leq\op{dim}(\rho_\mu)$,
\[
\sum_{g\in S_n}\rho_{\lambda}(g)[i,j] \overline{\rho_{\mu}(g)[k,l]} =
\left\{
\begin{array}{ll}
\frac{n!}{\op{dim}(\rho_\lambda)} & \mbox{if $\lambda = \mu$, $i=k$, and
$j = l$}\\
0 & \mbox{otherwise}
\end{array}
\right.
\]
\end{fact}

When $\lambda,\nu\vdash n$, we write $\chi_\lambda(\nu)$ to denote
the character $\chi_\lambda$ evaluated at an arbitrary $g\in C_\nu$,
and more generally if $f$ is a class function we write $f(\nu)$
to mean $f(g)$ for any $g\in C_\nu$.
\begin{fact}
\label{fact:character-order}
The sum of the squares of the characters of a conjugacy class over any
complete, irreducible set of representations of a group $G$ multiplied by the
order of the class is the order of $G$.
Thus, we have for $\lambda\vdash n$
\[
|C_\lambda|\sum_{\nu\vdash n} \chi_\nu(\lambda)^2 = n!
\]
\end{fact}

It will be necessary for us to be able to evaluate the characters associated
with the irreducible representations of the symmetric group in certain
instances.
The Murnaghan-Nakayama rule provides a tool for doing this---information on
the Murnaghan-Nakayama rule can be found in~\cite{Sternberg94}.


\section{Continuous-time quantum walks on $\Gamma(S_n,C_\lambda)$}

In this section we analyze the quantum walk on $\Gamma(S_n,C_\lambda)$
for $\lambda\vdash n$.
Our analysis implies that for some natural choices for $\lambda$
the quantum walk on $\Gamma(S_n,C_\lambda)$ does not have a uniform
limiting distribution with respect to the definition discussed in the
previous section.
In essence, the quantum walk has a significant ``blind spot'' consisting
of all $n$-cycles (i.e., permutations having cycle-structure
consisting of a single $n$-cycle).

This section is divided into three subsections.
First we prove a general result concerning the spectral decomposition
of quantum walks on $S_n$.
We then consider the case where the generating set consists of
the set of all transposition, and finally the case where the generating
set consists of all $p$-cycles for any choice of $p\in\{2,\ldots,n\}$.


\subsection{Spectral decomposition and periodicity}

Define $c_{\lambda}:S_n\rightarrow\mathbb{C}$ to be the unit vector
that is uniform on the conjugacy class $C_\lambda$ and zero everywhere else:
\[
c_\lambda[g] =\left\{
\begin{array}{ll}
\frac{1}{\sqrt{|C_{\lambda}|}} & \mbox{if $g\in C_{\lambda}$}\\
0 & \mbox{otherwise}.
\end{array}
\right.
\]

The analysis of quantum walks on $\Gamma(S_n,C_\lambda)$ is greatly
simplified by the fact that these walks are constant on conjugacy classes,
in the following sense.
\begin{prop}
Let $\alpha_t(g)$ denote the amplitude associated with vertex $g$ after
evolving the quantum walk on $\Gamma(S_n,C_{\lambda})$ for
time $t$, assuming the walk starts on a conjugacy class, i.e.,
$\alpha_t(g) = \left(U(t)c_{\lambda}\right)[g]$.
Then for all $t$, $\alpha_t$ is a class function.
\end{prop}

The following lemma will be one of the main tools used in our analysis.
\begin{theorem}
\label{theorem:spectrum}
Assume $H[g,h] = f(g^{-1}h)$ for all $g,h\in S_n$, where $f$ a class function
on $S_n$, and let $U(t) = e^{itH}$ for all $t\in\mathbb{R}$.
Then for any partitions $\lambda,\mu\vdash n$ we have
\[
c_{\lambda}^{\ast} U(t) c_{\mu} = 
\frac{\sqrt{|C_{\lambda}|} \sqrt{|C_{\mu}|}}{n!}
\sum_{\nu\vdash n}
\op{exp}\left(\frac{i t}{\op{dim}(\rho_\nu)}
\sum_{\gamma\vdash n} |C_\gamma| f(\gamma) \chi_\nu(\gamma) \right)
\chi_{\nu}(\lambda)
\chi_{\nu}(\mu).
\]
\end{theorem}
In order to prove this theorem we will use the following lemma,
by which a complete orthogonal set of eigenvectors and eigenvalues
of $U(t)$ can be obtained.

\begin{lemma}
\label{lemma:walk-spectrum}
Assume $H[g,h] = f(g^{-1}h)$ for all $g,h\in S_n$, where $f$ is a class
function on $S_n$.
Define vectors $\psi_{\nu,i,j}:S_n \rightarrow \mathbb{C}$ for each
$\nu\vdash n$, $1\leq i,j\leq \op{dim}(\rho_{\nu})$, as follows:
\[
\psi_{\nu,i,j}[g] = \rho_{\nu}(g)[i,j]
\]
for all $g\in S_n$.
Then each $\psi_{\nu,i,j}$ is an eigenvector of $H$ with associated eigenvalue
\[
\frac{1}{\op{dim}(\rho_\nu)}\sum_{\gamma\vdash n} |C_\gamma| f(\gamma)
\chi_\nu(\gamma).
\]
Moreover, these eigenvectors are pairwise orthogonal and span the entire
space $\mathbb{C}^{S_n}$.
\end{lemma}

\noindent {\bf Remark.}
The fact described in Lemma~\ref{lemma:walk-spectrum} is not new---for
instance, it is discussed in Section 3E of \cite{Diaconis88} for general
finite groups.
A short proof of the lemma follows.
\vspace{2mm}

\noindent {\bf Proof of Lemma~\ref{lemma:walk-spectrum}.}
For each $g\in S_n$ we have
\[
(H\psi_{\nu,i,j})[g] =
\sum_{h\in G}
H[g,h]\psi_{\nu,i,j}[h] 
= 
\sum_{h\in G}
f(g^{-1}h)
\rho_\nu(h)[i,j] 
= 
\sum_{h\in G}
f(h)
\rho_\nu(gh)[i,j].
\]
Now, since $\rho_\nu$ is a homomorphism, we have
$\rho_\nu(gh)=\rho_\nu(g)\rho_\nu(h)$, which implies
\[
(H\psi_{\nu,i,j})[g] =
\sum_{k=1}^{\op{dim}(\rho_\nu)}
\rho_\nu(g)[i,k]
\left(\sum_{h\in S_n}f(h) \rho_\nu(h)\right)[k,j]
=
\sum_{k=1}^{\op{dim}(\rho_\nu)}
\rho_\nu(g)[i,k]
\hat{f}(\rho_\nu)[k,j].
\]
By Fact~\ref{fact:Fourier-class} we see that
\[
(H\psi_{\nu,i,j})[g] =
\frac{1}{\op{dim}(\rho_\nu)}\sum_{h\in S_n} f(h) \chi_\nu(h) \rho_\nu(g)[i,j]
= \left(
\frac{1}{\op{dim}(\rho_\nu)}\sum_{\gamma \vdash n}|C_{\gamma}| f(\gamma)
\chi_\nu(\gamma) \right)
\psi_{\nu,i,j}[g].
\]
This establishes that the vectors $\psi_{\nu,i,j}$ are eigenvectors with
associated eigenvalues as claimed.
The fact that these eigenvectors are pairwise orthogonal follows from
Fact~\ref{fact:orthogonality-relations} and the fact that they span the
entire space $\mathbb{C}^{S_n}$ follows from this orthogonality along
with Fact~\ref{fact:character-order}.
\qed

\noindent
{\bf Proof of Theorem~\ref{theorem:spectrum}.}
By Lemma~\ref{lemma:walk-spectrum} we may write
\[
H = \sum_{\nu,j,k}
\left(
\frac{1}{\op{dim}(\rho_\nu)}
\sum_\gamma |C_\gamma| f(\gamma) \chi_\nu(\gamma)\right)
\frac{\psi_{\nu,j,k}\psi_{\nu,j,k}^{\ast}}
{\|\psi_{\nu,j,k}\|^2}
\]
and therefore
\[
U(t) = \sum_{\nu,j,k}
\op{exp}\left(\frac{i t}{\op{dim}(\rho_\nu)}
\sum_\gamma |C_\gamma| f(\gamma) \chi_\nu(\gamma)\right)
\frac{\psi_{\nu,j,k}\psi_{\nu,j,k}^{\ast}}
{\|\psi_{\nu,j,k}\|^2}
\]
Let $X_{\lambda}:S_n\rightarrow\mathbb{C}$ denote the characteristic
function of $C_{\lambda}$ for $\lambda \vdash n$.
Then we have that
\[
c_{\lambda}^{\ast}\psi_{\nu,j,k}
=
\frac{1}{\sqrt{|C_{\lambda}|}}
\sum_{g\in S_n}
X_{\lambda}(g)\rho_{\nu}(g)[j,k]
=
\frac{1}{\sqrt{|C_{\lambda}|}}
\hat{X_{\lambda}}(\rho_{\nu})[j,k]
=
\left\{
\begin{array}{ll}
\frac{\sqrt{|C_{\lambda}|} \,\chi_{\nu}(\lambda)}{\op{dim}(\rho_{\nu})}
& \mbox{if $j = k$}\\
0 & \mbox{otherwise}
\end{array}
\right.
\]
by Fact~\ref{fact:Fourier-class}.
By Fact~\ref{fact:orthogonality-relations} we have
$\|\psi_{\nu,j,k}\|^2 = \frac{n!}{\op{dim}(\rho_\nu)}$.
So,
\begin{eqnarray*}
c_{\lambda}^{\ast} U(t) c_{\mu}
& = &
\frac{1}{n!}
\sum_{\nu}
\op{exp}\left(\frac{i t}{\op{dim}(\rho_\nu)}
\sum_{\gamma\vdash n} |C_\gamma| f(\gamma) \chi_\nu(\gamma)\right)
\op{dim}(\rho_{\nu})
\sum_{1\leq j,k\leq \op{dim}(\rho_{\nu})}
c_{\lambda}^{\ast}\psi_{\nu,j,k}\,
\psi_{\nu,j,k}^{\ast}c_{\mu}\\
& = &
\frac{\sqrt{|C_{\lambda}|} \sqrt{|C_{\mu}|}}{n!}
\sum_{\nu\vdash n}
\op{exp}\left(\frac{i t}{\op{dim}(\rho_\nu)}
\sum_{\gamma\vdash n} |C_\gamma| f(\gamma) \chi_\nu(\gamma)\right)
\chi_{\nu}(\lambda)\chi_{\nu}(\mu),
\end{eqnarray*}
which is what we wanted to show.
\qed

Theorem~\ref{theorem:spectrum} implies the following interesting fact.

\begin{prop}
\label{prop:periodicity}
Any continuous-time quantum walk on the Cayley graph of the symmetric group
for which the generators form conjugacy classes is periodic, with period
$2\pi/k$ for some $k\in\{1,2,3,\ldots\}$.
\end{prop}

\noindent
{\bf Proof.}
Using Fact~\ref{fact:Fourier-class} we see that the quantity
$|C_\gamma|\chi_\nu (\gamma)/\op{dim}(\rho_\nu)$ is a sum of matrix 
elements of irreducible representations.
This quantity is independent of the particular choice of the basis for
the irreducible representations, so we may choose that basis that corresponds
to Young's natural form, in which all of the matrix entries are integer
valued, implying that the quantity itself is integer valued.
Using Fact~\ref{fact:character-order} and
Theorem~\ref{theorem:spectrum} therefore have that
$U(2\pi)=U(0)=I$.
Thus the period of the walk must divide $2\pi$.
\qed

We have not discussed mixing times in this paper, but the previous proposition
implies that quantum walks on Cayley graphs of $S_n$ reach their limiting
distribution quickly, and when calculating the limiting distribution it is only
necessary to average over times in the range $[0,2\pi]$.
Note that in terms of implementation, this does not mean that the walk mixes
in constant time; some number of operations that is polynomial in the
degree of the graph and in some accuracy parameter is required to implement
such a walk, assuming the ability to compute the neighbors of each vertex.
See \cite{AharonovT03,ChildsC+03} for further details.


\subsection{Cayley graphs of $S_n$ generated by transpositions}

For the Cayley graph of $S_n$ generated by the transpositions,
Theorem~\ref{theorem:spectrum} has various implications that we
discuss in this section.
We will require explicit values for various characters of the symmetric group,
which we now mention.
Using the Murnaghan-Nakayama rule it can be shown that
\[
\chi_\nu \left((n)\right) 
= \left\{
\begin{array}{ll}
(-1)^{n-k} & \mbox{for $\nu = (k,1,\ldots,1)$, $k \in \{1,\ldots,n\} $ }\\
\hspace{3mm} 0 & \mbox{otherwise}
\end{array}
\right.
\]
and 
\[
\chi_{(k,1,\ldots,1)}\left(\mathrm{id} \right) 
= \op{dim}(\rho_{(k,1,\ldots,1)})
= \binom{n-1}{k-1}.
\]
For the characters at the transpositions, it is known \cite{Ingram50} that
\[
\chi_\nu(\tau) = \frac{\op{dim}(\rho_\nu)}{\binom{n}{2}}
\sum_j \left(\binom{\nu_j}{2} - \binom{\nu_j'}{2}\right).
\]
Here, $\tau$ is any transposition, $\nu'$ is the partition generated by
transposing the Young diagram of $\nu$, while $\nu_j$ and $\nu_j'$ are the
$j^{th}$ components of the partitions $\nu$ and $\nu'$.

Substituting these values into Theorem~\ref{theorem:spectrum} gives
\[
c_{\lambda}^{\ast} U(t) c_{\mu} = 
\frac{\sqrt{|C_{\lambda}|} \sqrt{|C_{\mu}|}}{n!}
\sum_{\nu\vdash n}
\op{exp}\left(i t
\sum_j \left(\binom{\nu_j}{2} - \binom{\nu_j'}{2}\right)\right)
\chi_{\nu}(\lambda)
\chi_{\nu}(\mu)
\]
for the quantum walk on $\Gamma(S_n,C_{(2,1,\ldots,2)})$, and
specifically for the case where $\mu = (1,\ldots,1)$ and
$\lambda = (n)$ it follows that
\begin{eqnarray*}
c_{(n)}^{\ast} U(t) c_{(1,\ldots,1)} & = &
\frac{1}{\sqrt{n\cdot n!}}\sum_{k=1}^n\op{exp}\left(it\left(
\binom{k}{2} - \binom{n-k+1}{2}\right)\right)(-1)^{n-k}\binom{n-1}{k-1}\\
& = &
\frac{\left(2 i \sin (t n/2)\right)^{n-1}}{\sqrt{n\cdot n!}}.
\end{eqnarray*}
In particular,
\begin{equation}
\label{eq:max_prob}
\max_t \left|c_{(n)}^{\ast} U(t) c_{(1,\ldots,1)}\right|^2 
= \frac{2^{2n-2}}{n\cdot n!},
\end{equation}
where the maximum occurs for $t = (2k+1)\pi/n$, $k\in\mathbb{Z}$.

Eq.~\ref{eq:max_prob} has the following interpretation.
If we start a quantum walk on $\Gamma(S_n,C_{(2,1,\ldots,1)})$ at the identity
element and evolve for any amount of time and measure, the probability
to measure some $n$-cycle is at most $\frac{2^{2n-2}}{n\cdot n!}$
as opposed to probability approaching $\frac{1}{n}$ for the classical case.
The probability to measure any particular $n$-cycle is therefore at most
$\frac{2^{2n-2}}{(n!)^2}$, as opposed to some number approaching
$\frac{1}{n!}$ classically.
The probabilities in the quantum case are smaller by a factor that
is exponential in $n$.

As discussed in Section~\ref{sec:walks-definition}, we will 
denote by $P_t$ the distribution on $S_n$ obtained by performing the quantum
walk on $\Gamma(S_n,C_{(2,1,\ldots,1)})$ for time $t$ starting at the identity
then measuring.
The above analysis gives a lower bound for the total variation distance
of $P_t$ from the uniform distribution:
\[
\|P_t - \mathrm{uniform}\| \geq \frac{1}{n} - \frac{2^{2n-2}}{n\cdot n!}
\]
for all values of $t$.
This bound follows from considering only the $n$-cycles, and we
believe the true bound to be much larger.
Numerical simulations support this claim, but thus far we only have exact
expressions for the $n$-cycles.

Given that we have an exact expression for the probability $P_t[g]$
for any $n$-cycle $g$, it is easy to determine the probability
associated with any $n$-cycle in the limiting distribution.
To be precise, let
\[
\bar{P}[g] = \frac{1}{2\pi}\int_0^{2\pi}P_t[g] \,dt
\]
for each $g\in S_n$.
Then for any $g\in C_{(n)}$ we have
\[
\bar{P}[g] = \frac{1}{2\pi}\int_0^{2\pi}
\frac{\left(2 \sin (t n/2)\right)^{2n-2}}{(n!)^2}dt
=
\frac{\binom{2n-2}{n-1}}{(n!)^2}.
\]
Somewhat surprisingly, this average probability associated with reaching a
given $n$-cycle is not unique to the particular choice of $C_{(2,1,\ldots,1)}$
as a generating set, as shown in the next subsection.


\subsection{Other generating sets}

We have not been able to obtain tractable expressions for the amplitudes
associated with quantum walks for other generating sets besides
$C_{(2,1,\ldots,1)}$.
However, we can prove some facts concerning the limiting distributions for
such walks in the case that the generating set consists of all
$p$-cycles for any choice of $p$.
(In case $p$ is odd, we must keep in mind that only the alternating
group is being generated.)
Again we will focus on the probability of reaching $n$-cycles starting
from the identity.

Consider the quantum walk on $\Gamma(S_n,C_{\gamma})$, where $\gamma$ is
any partition.
According to Theorem~\ref{theorem:spectrum}, the probability associated with a
given conjugacy class $C_{\lambda}$ when starting from a uniform superposition
on another class $C_{\mu}$ after time $t$ is given by
\[
|c_\lambda^* U(t)c_\mu|^2 = 
\frac{|C_\lambda||C_\mu|}{(n!)^2} \sum_{\nu,\eta} 
\exp\left(it \, |C_\gamma|\left(\frac{\chi_\nu(\gamma)}{\op{dim}(\rho_\nu)} 
- \frac{\chi_\eta(\gamma)}{\op{dim}(\rho_\eta)}\right) \right)
\chi_\nu(\lambda)\chi_\nu(\mu)\chi_\eta(\lambda)\chi_\eta(\mu).
\]
As before, we let $\bar{P}$ denote the limiting distribution of the walk when
starting from the identity.
Since our walks are periodic with period $2\pi$, we therefore have
\begin{eqnarray*}
\bar{P}[g] & = & \frac{1}{(n!)^2}
\frac{1}{2\pi}\int_0^{2\pi}
\sum_{\nu,\eta} 
\exp\left(it \, |C_\gamma|\left(\frac{\chi_\nu(\gamma)}{\op{dim}(\rho_\nu)} 
- \frac{\chi_\eta(\gamma)}{\op{dim}(\rho_\eta)}\right) \right)
\chi_\nu(g)\op{dim}(\rho_\nu)\chi_\eta(g)\op{dim}(\rho_\eta)dt\\
& = & 
\frac{1}{(n!)^2} \sum_{\nu,\eta}^{\ast} 
\chi_\nu(g)\op{dim}(\rho_\nu)\chi_\eta(g)\op{dim}(\rho_\eta).
\end{eqnarray*}
Here the asterisk denotes that the sum is over all partitions
$\nu,\eta$ subject to the condition
\begin{equation}
\label{eq:same-fraction}
\frac{\chi_\nu(\gamma)}{\op{dim}(\rho_\nu)} = 
\frac{\chi_\eta(\gamma)}{\op{dim}(\rho_\eta)}.
\end{equation}
Observe that the choice of generators only affects the average distribution by
determining what values other than $\nu = \eta$ are included in the sum.
More generally, the average probability associated with obtaining some
element in $C_{\lambda}$ when starting the walk on the uniform superposition
over $C_{\mu}$ is given by
\[
\frac{|C_\lambda||C_\mu|}{(n!)^2} \sum_{\nu,\eta}^{\ast} 
\chi_\nu(\lambda)\chi_\nu(\mu)\chi_\eta(\lambda)\chi_\eta(\mu).
\]

For the remainder of this section, let $\gamma\vdash n$ be the partition
corresponding to the $p$-cycles, i.e., $\gamma = (p,1,\ldots,1)$, where
$2\leq p\leq n$.
We will consider the limiting distribution for the walk on
$\Gamma(S_n,C_{\gamma})$.
The Murnaghan-Nakayama rule can be used to give the
following formula for some of the character values at the $p$-cycles
for $1\leq p \leq n-1$:
\[
\chi_{(k,1,\ldots,1)}((p,1,\ldots,1)) = 
\binom{n-p-1}{k-p-1} + (-1)^{p+1}\binom{n-p-1}{k-1}.
\]

First, let us assume that $p$ is even, so $C_{\gamma}$ generates the
entire symmetric group.
We will consider first the case $p\leq\lceil\frac{n}{2}\rceil$.
To determine the affect of the generators on the walk we need to determine
when Eq.~\ref{eq:same-fraction} holds.
In the particular case where we are interested in $\bar{P}[g]$ for $g$ 
an arbitrary $n$-cycle, we may take advantage of the fact that
$\chi_{\nu}(g) = 0$ unless $\nu = (k,1,\ldots,1)$ for some $k$.
Thus, it will be sufficient to determine for what values of $k$ and $k'$ we
have
\begin{equation}
\label{eq:k-and-k'}
\frac{\chi_{(k,1,\ldots,1)}(\gamma)}{\op{dim}(\rho_{(k,1,\ldots,1)})} = 
\frac{\chi_{(k',1,\ldots,1)}(\gamma)}{\op{dim}(\rho_{(k',1,\ldots,1)})}.
\end{equation}
Define
\[
f(k) = \frac{\chi_{(k,1,\ldots,1)}(\gamma)}{\op{dim}(\rho_{(k,1,\ldots,1)})}.
\]
Since $p$ is even we have
\[
f(k) = \frac{1}{\binom{n-1}{k-1}}\left(
\binom{n-p-1}{k-p-1} - \binom{n-p-1}{k-1} \right)
=
\frac{1}{\binom{n-1}{p}}
\left(\binom{k-1}{p} - \binom{n-k}{p} \right),
\]
and thus
\[
f(k+1)-f(k) = \frac{1}{\binom{n-1}{p}}
\left(\binom{k-1}{p-1} + \binom{n-k-1}{p-1}\right).
\]
For $2\leq p \leq \lceil \frac{n}{2} \rceil$ this difference is positive,
so $f(k)$ is strictly increasing and is therefore 1-to-1.
Thus, Eq.~\ref{eq:k-and-k'} is satisfied only when $k = k'$ and so
the average probability of reaching a given $n$-cycle is
\[
\frac{1}{(n!)^2} \sum_{k=1}^n \chi_{(k,1,\ldots,1)}((n))^2 
\chi_{(k,1,\ldots,1)}(\mathrm{id})^2 
= \frac{1}{(n!)^2} \sum_{k=1}^n \binom{n-1}{k-1}^2
=\frac{1}{(n!)^2}  \binom{2n-2}{n-1}.
\]
The implication is that in this case the probability associated with any
$n$-cycle in the limiting distribution is identical to the walk where the
generating set consists of the transpositions.

For $\lceil \frac{n}{2} \rceil<p\leq n-1$, the situation is slightly more
complicated.
Since $f(k+1) - f(k)$ is still nonnegative, $f(k)$ is increasing, but
not strictly:
\[
f(k) = f(k')\;\Leftrightarrow\;k,k'\in\{n-p+1,\ldots,p\}
\;\mbox{or}\;k = k'.
\]
The average probability to reach a given $n$-cycle is therefore
\begin{eqnarray*}
\lefteqn{\frac{1}{(n!)^2} \sum_{k=1}^n \binom{n-1}{k-1}^2
+
\frac{1}{(n!)^2}
\sum_{\stackrel{\scriptstyle n-p+1\leq k,k'\leq p}{\scriptstyle k\not=k'}}
(-1)^{k+k'}\binom{n-1}{k-1}\binom{n-1}{k'-1}}\\
& = &
\frac{1}{(n!)^2} \left(
\sum_{k=1}^{n-p} \binom{n-1}{k-1}^2 +
\sum_{k=p+1}^{n} \binom{n-1}{k-1}^2\right)+
\frac{1}{(n!)^2}\left(\sum_{n-p+1\leq k \leq p}(-1)^k\binom{n-1}{k-1}\right)^2.
\end{eqnarray*}
If $n$ is even we have
\[
\sum_{n-p+1\leq k \leq p}(-1)^k\binom{n-1}{k-1}=0
\]
and thus the average probability to reach a given $n$-cycle
can be rewritten as
\[
\frac{1}{(n!)^2} \left(
\sum_{k=1}^{n-p} \binom{n-1}{k-1}^2 +
\sum_{k=p+1}^{n} \binom{n-1}{k-1}^2\right) = 
\frac{2}{(n!)^2}\sum_{k=1}^{n-p} \binom{n-1}{k-1}^2.
\]
This probability is therefore smaller than for the case
$p\leq\lceil\frac{n}{2}\rceil$.
For $n$ odd we have
\[
\sum_{n-p+1\leq k \leq p}(-1)^k\binom{n-1}{k-1} = 2\binom{n-2}{p-1},
\]
so the average probability to reach a given $n$-cycle is
\[
\frac{1}{(n!)^2} \left(
2\sum_{k=1}^{n-p} \binom{n-1}{k-1}^2 +
4\binom{n-2}{p-1}^2\right).
\]
This probability is also smaller than the average probability of reaching
an $n$-cycle for the case $p\leq \lceil n/2\rceil$ since
\[
\sum_{k=n-p+1}^p\binom{n-1}{k-1}^2 \geq
(2p-n-1)\binom{n-1}{p-1}^2 \geq 4\binom{n-2}{p-1}^2.
\]
Finally, for the special case where $p=n$ (still assuming $p$ is even),
we again have the situation that Eq.~\ref{eq:k-and-k'} is satisfied only
when $k = k'$, and thus the average probability to reach an $n$-cycle is
again $\frac{1}{(n!)^2}\binom{2n-2}{n-1}$.

Now we consider the case where $p$ is odd.
In this case, the quantum walk is taking place on the alternating
group.
If $n$ is even, there is of course therefore zero probability to reach any
given $n$-cycle, so assume $n$ is odd.
Letting $f(k)$ be as defined before and considering that $p$ is odd, we have
\[
f(k) = \frac{1}{\binom{n-1}{p}}\left(\binom{k-1}{p} + \binom{n-k}{p}\right),
\]
which is symmetric about $\frac{n+1}{2}$.
By a similar argument to before, it can be shown that
in the case $p\leq \frac{n+1}{2}$, the function $f(k)$ is
strictly decreasing for $1\leq k \leq \frac{n+1}{2}$ and
strictly increasing for $\frac{n+1}{2}\leq k \leq n$.
Eq.~\ref{eq:k-and-k'} is therefore satisfied if and only if
$k = k'$ or $k = n - k' + 1$.
This gives an average probability of
\[
\frac{2}{(n!)^2}\binom{2n-2}{n-1} - 
\frac{1}{(n!)^2}\binom{n-1}{\frac{n-1}{2}}^2
\]
to reach a given $n$-cycle.
This is slightly less than twice the probability of reaching a given
$n$-cycle for the case $p$ even and $p\leq \lceil\frac{n}{2}\rceil$,
which is not surprising since the alternating group is half the size of
the symmetric group.

Next suppose $p>\frac{n+1}{2}$.
Then in the range $1\leq k\leq n-p+1$ the function $f(k)$ is strictly
decreasing, in the range $n-p+1\leq k\leq p$ the function $f(k)$ is constant
($f(k) = 0$ in this range), and in the range $p\leq k\leq n$ the function
$f(k)$ is strictly increasing.
This implies that the average probability of reaching each $n$-cycle is
\[
\frac{1}{(n!)^2}
\left(4\sum_{k = 1}^{n-p}\binom{n-1}{k-1}^2 +
4\binom{n-2}{p-1}^2\right).
\]
Lastly, in the case $p=n$, we are back to the situation that
$f(k) = f(k')$ if and only if $k = k'$ or $k = n - k' + 1$, giving an average
probability of
\[
\frac{2}{(n!)^2}\binom{2n-2}{n-1} - 
\frac{1}{(n!)^2}\binom{n-1}{\frac{n-1}{2}}^2
\]
to reach each $n$-cycle.

We summarize the facts we have just discussed in the following table.
\begin{center}
\renewcommand{\arraystretch}{1.5}
\begin{tabular}{|c|c|c|c|}
\hline
Parity of $n$ & Parity of $p$ & Range of $p$ &
Average probability to reach each $n$-cycle\\
\hline
\hline
even or odd & even & $2\leq p \leq \lceil\frac{n}{2}\rceil$ &
$\frac{1}{(n!)^2}\binom{2n-2}{n-1}$\\
\hline
even & even & $\frac{n}{2} + 1 \leq p \leq n-1$ &
$\frac{2}{(n!)^2}\sum_{k=1}^{n-p}\binom{n-1}{k-1}^2$\\
\hline
odd & even  & $\frac{n+1}{2} + 1 \leq p \leq n-1$ &
$\frac{2}{(n!)^2}\sum_{k=1}^{n-p}\binom{n-1}{k-1}^2
+ \frac{4}{(n!)^2}\binom{n-2}{p-1}^2$\\
\hline
even & even & $p=n$ &
$\frac{1}{(n!)^2}\binom{2n-2}{n-1}$\\
\hline
even & odd  & --- & 0\\
\hline
odd & odd  & $2\leq p \leq \frac{n+1}{2}$ &
$\frac{2}{(n!)^2}\binom{2n-2}{n-1}
-\frac{1}{(n!)^2}\binom{n-1}{\frac{n-1}{2}}^2$\\
\hline
odd & odd  & $\frac{n+1}{2} + 1 \leq p \leq n-1$ &
$\frac{4}{(n!)^2}\sum_{k=1}^{n-p}\binom{n-1}{k-1}^2
+ \frac{4}{(n!)^2}\binom{n-2}{p-1}^2$\\
\hline
odd & odd & $p=n$ &
$\frac{2}{(n!)^2}\binom{2n-2}{n-1}
-\frac{1}{(n!)^2}\binom{n-1}{\frac{n-1}{2}}^2$\\
\hline
\end{tabular}
\end{center}
\vspace{2mm}

We have the following lower bounds on the total variation distance of the
limiting distribution from the uniform distribution.
As for the case of the case of the quantum walk generated by the
transpositions, this bound follows just from considering the $n$-cycles,
and we believe the true distance from uniform to be much larger.
\begin{itemize}
\item
Let $p\in\{2,\ldots,n\}$ be even, let $\gamma = (p,1,\ldots,1) \vdash n$ and
let $\bar{P}$ denote the limiting distribution of the quantum walk on
$\Gamma(S_n,C_{\gamma})$.
Then
\[
\|\bar{P} - \mathrm{uniform}(S_n)\| \geq 
\frac{1}{n} - \frac{1}{n\cdot n!}\binom{2n-2}{n-1}.
\]

\item
Let $n$ be odd, let $p\in\{2,\ldots,n\}$ be odd, let
$\gamma = (p,1,\ldots,1) \vdash n$ and let $\bar{P}$ denote the limiting
distribution of the quantum walk on $\Gamma(S_n,C_{\gamma})$.
Then
\[
\|\bar{P} - \mathrm{uniform}(A_n)\| \geq \frac{2}{n} - \frac{2}{n\cdot n!}
\binom{2n-2}{n-1} + \frac{1}{n\cdot n!}\binom{n-1}{\frac{n-1}{2}}^2.
\]
\end{itemize}


\section{Conclusion}

In this paper we have studied some of the properties of continuous-time
quantum walks on Cayley graphs of the symmetric group.
Many questions concerning these walks remain unanswered.
One obvious question that we have not attempted to address in this paper
is whether quantum walks on the symmetric group can be applied in the context
of quantum algorithms.
In terms of specific properties of these walks, we have focused on
the limiting distribution---is the limiting distribution bounded away from
uniform by a constant?
Many other properties of these walks may be of interest as well.
For instance, the effect of decoherence on these walks is an interesting
topic to consider.


\bibliographystyle{plain}



\end{document}